\documentstyle[aps,preprint,prl]{revtex}  
\def\({\begin{equation}}
\def\){\end{equation}}
\begin{document}                
\title{On the Problem of Spin Diffusion in 1D Antiferromagnets}
\author{B. N. Narozhny}
\address{Department of Physics, Rutgers University, Piscataway, NJ 08855} 
\maketitle
\begin{abstract}  
We study the problem of spin diffusion in magnetic systems without 
long-range order. We discuss the example of the 1D spin chain. For 
the system described by the Heisenberg Hamiltonian we show that
there are no diffusive excitations. However, the addition of an 
arbitrarily small dissipation term, such as the spin-phonon 
interaction leads to diffusive excitations in the long 
time limit. For those excitations we estimate the spin-diffusion 
coefficient by means of the renormalisation group analysis.              

\end{abstract}

\narrowtext
\section{Introduction}

\ 

The spin dynamics in magnetic systems without long-range order is a long-standing problem. 
It is generally believed \cite{ghar,bouc,luri} that 
in the high temperature limit, where no long-range order is present, the microscopic spin 
fluctuations are governed by the diffusion equation, i.e. that at small frequencies 
and momenta the spin correlation function has a diffusive pole

\  

\begin{equation}
< S S > =g(\omega, k)={Z\over{i\omega + Dk^2}}
\label{corrfunc}
\end{equation}

\  

\noindent
where $D$ is the diffusion constant and Z is the residue.

Although this idea of spin diffusion is quite common, we are not 
aware of any theoretical approach, within which one could "derive" the correlation function 
Eq.\ (\ref{corrfunc}), starting from the nearest neighbour Heisenberg Hamiltonian

\  

\begin{equation}
H_H=\sum_i \;  J_{i}\;  \vec S_i\cdot\vec S_{i+1}.
\label{heisham}
\end{equation}

\  

In this paper we are presenting such an approach for 1D spin-1/2 Heisencberg chains of the infinite length. 
At any non-zero temperature the chain is in disordered state with exponentially decaying spin-spin correlations
and even at $T=0$ the correlations decay as a power law, so there is no true long-range order.
We show that in one dimension it is impossible to derive 
the diffusive form of the spin-spin correlation function Eq.\ (\ref{corrfunc}) from the Heisenberg Hamiltonian
Eq.\ (\ref{heisham}), without any kind of additional dissipation mechanism.  
We also present a general argument supporting this result. We show then, that
if one takes into account an additional dissipation, for example due to spin-phonon interaction (which is always
present in any real system at finite temperature), then the renormalisation group approach leads 
to the correlation function Eq.\ (\ref{corrfunc}) in the long-time asymptotic.

Our results could be applied to the materials  like $KCuF_3 , CuSO_4\cdot5H_2O , Sr_2CuO_3 $. 
In a broad temperature range they are nearly ideal 1D antiferromagnetic chains with the coupling constant $J$ ranging 
from 1.45 K in 
$CuSO_4\cdot5H_2O$ \cite{ghar} to 190 K in $KCuF_3$ \cite{yama,ueda} and 1300 K in the novel $Sr_2CuO_3$ \cite{ami}.  
The diffusion equation was successfully used in a number of 
experimental papers to fit the data and explain the results of the experiments \cite{ghar,bouc}. This approach was also 
confirmed by computer simulations \cite{luri}.

The rest of this paper is organized as follows. In Section II we briefly review the mapping of the spin system 
onto 1D fermions. In the Section III we map the fermions onto a boson system, justifying bosonisation 
procedure at finite temperatures. Section IV gives the calculation, which shows the absence of spin diffusion in the 
Heisenberg model Eq.\ (\ref{heisham}) and presents some general qualitative arguments, supporting this conclusion,
based on the theory of the Sine-Gordon
equation. Section V discusses the spin-phonon interaction and its renormalisation.
Section VI is conclusions and discussion of results. The details of the calculations are presented in Appendices.

\ 

\section{Fermion Model}

\ 

In this section we review the well-known procedure of mapping the Heisenberg model Eq.\ (\ref{heisham}) on a
system of spinless fermions \cite{jor} and establish the notations.

The spin model Eq.\ (\ref{heisham}) can be transformed into a model of spinless fermions, noting that 
operators $S_i^{+}$ and $S_i^{-}$ anticommute. The Jordan-Wigner transformation then relates spin to fermion
operators $(a_i^{\;} ,a_i^{\dag})$ via

\begin{eqnarray}
&&S_i^{+}=a_i^{\dag} exp \left(i\pi \sum_{j=1}^{i-1}\; a_j^{\dag} a_j^{\;} \right) , \nonumber \\
&&S_i^z= a_i^{\dag} a_i^{\;} - {1\over2}
\label{jordwig}
\end{eqnarray}

\  

\noindent
When transforming the Hamiltonian Eq.\ (\ref{heisham}) the spin-flip terms give rise to the motion of the 
fermions (kinetic energy in the fermion Hamiltonian) and $S^z_i \cdot S^z_{i+1}$ interaction leads to a fermion-fermion
interaction between adjacent cites. Since $S^z$ is quadratic in fermion operators, the interaction between 
fermions is the four-particle interaction. Since the original spin model was formulated on a lattice, all 
possible types of four-fermion interaction are present, including the umklapp term. 

The 1D fermion models are often treated using bosonisation \cite{shan}. In the case of massless fermions with the
four-fermion interaction with small momentum transfer it allows exact solution.
It is shown, that in the thermodynamic limit that 
system can be mapped onto a system of free bosons. The propagator of a free boson has a pole at $\omega = ck$
and therefore corresponds to the particle, propagating without dissipation, 
so that this interaction can not lead to any significant change in the
long-time dynamics of the system. 
Therefore the only source of dissipation might be interaction with large momentum transfer, such as the umklapp term.

The treatment of the umklapp term is much more complicated because in 
the boson language it corresponds to the highly non-linear term, namely $\cos(2\beta\phi)$ (where
$\phi(x,t)$ is the boson field and the number $\beta$ is a parameter of the transformation, depending on the
fermion interaction with small momentum transfer). This term does not conserve momentum and is only marginally irrelevant, so
at $T>0$ it might lead to some new dynamics. Thus we should investigate the 
impact of that term on the long-time fermion (and thus spin) dynamics, disregarding all other possible 
four-fermion interaction terms. The next two Sections present the results of this investigation. 

Since we are interested only in the low-energy behaviour of the system, we can linearize the fermion kinetic
energy, thus allowing exact bosonisation. Therefore our model Hamiltonian is 

\  

\begin{equation}
H_m=v_F\sum_{k} \left[ (k-k_F) \psi_R^{\dag}\psi_R^{\;}  + (-k-k_F) \psi_L^{\dag} \psi_L^{\;} \right]
+V\sum\left(\psi_R^{\;}\psi_R^{\;}\psi_L^{\dag}\psi_L^{\dag} + h.c. \right),
\label{fermham}
\end{equation}

\  

\noindent
where $\psi_{R(L)}$ is the operator of a "right" ("left") mover.  The quantity we are interested in here is 
the density-density correlation function, which corresponds to the $ \langle S^z_i \cdot S^z_j \rangle $ correlator in the spin problem.
We shall now investigate, whether the long-time asymptotic of this correlator has the diffusive 
form, Eq.\ (\ref{corrfunc}).

\ 

\section{Bosonisation}

\ 

\ 

As we mentioned above our main technical tool will be bosonisation. The procedure is well established in 1D at zero
temperature. We consider finite temperatures and we are looking 
for the long-time asymptotic of the spin-spin correlator. It is quite difficult to obtain the results in that 
limit in the Matsubara technique, since the analytic continuation from the Matsubara frequencies $\omega_n$ to
real frequencies much less that the inverse temperature would require a precise knowledge of the Green's 
functions on the infinite range of $\omega_n$, which is usually not the case in perturbation theory. Therefore we
have to resort to the Keldysh technique \cite{keld}, which incorporates finite temperatures and real-time representation. In
this Section we construct the bosonisation procedure for the Keldysh technique and then confirm it's correctness
by comparing the results of the perturbation theory for bosons and fermions. The fermion and boson Green's 
functions in space-time representation, which we are using in our calculations, are presented in Appendix A.

Thirty years ago Keldysh has presented a field theoretical technique to calculate the real-time correlation
functions of a quantum system. To allow the treatment of advanced and retarded correlators, Keldysh introduced
the time contour C (Fig. 1) with the upper branch going in positive direction from $-\infty$ to $+\infty$ and the
backward lower branch. All the operator products are now time-ordered along the contour C. To distinguish
particles on the upper and the lower branches of the contour the fermion field operator $\psi_\gamma$ is given
an index $\gamma$, which equals to $1$ on the upper branch and $2$ on the lower. Green's functions become 
$2\times 2$ matrices with respect that index.

In a one dimensional fermion problem we have four different operators $\psi_\gamma^{L(R)}$ - left and right movers on 
both branches of the contour. Since operators on each separate branch are completely analogues to the 
zero-temperature operators, whe can proceed with the bosonisation separately on each branch in exactly the same
way as at $T=0$. Thus we introduce two boson fields, $\phi_\gamma$, (one for each branch), which we shall treat as two
components of the Keldysh field. The resulting bosonised Hamiltonian will thus be formulated in the Keldysh
technique also.

The left $\phi_\gamma^L$ and right $\phi_\gamma^R$ moving bose fields expressed via $\phi_\gamma$ and its canonically 
conjugate $P_\gamma$.

\

\begin{eqnarray}
\phi_\gamma^{L(R)}(x)  && = {1\over 2} \Bigg[\phi_\gamma(x) \mp \int_{-\infty }^{x} P_\gamma(x') dx'\Bigg]
\nonumber\\
&& \;
\nonumber\\
&& = \pm \int_0^{\infty } { {dp}\over{2\pi\sqrt{2|p|}} } e^{\displaystyle -\alpha |p|}
\Bigg[\phi_\gamma(p) e^{ipx} + h.c. \Bigg].
\label{boss}
\end{eqnarray}

\ 

\ 

\noindent
As in the usual procedure $\phi_\gamma^{L(R)}$ are functions of only
$(x\mp t)$. 

The fermion operators are constructed in analogy with the zero temperature case

\

\begin{equation}
\psi^{L(R)}_\gamma \sim {1\over\sqrt{\alpha}} \exp \bigg(\pm i\beta{\;}\phi_\gamma^{L(R)}\bigg)
\label{opereq}
\end{equation}

\ 

\noindent
where $\beta^2= 4\pi$ and the upper sign corresponds to the left mover.

The commutation relations between fermion fields hold by exactly the same reason \cite{man} as at $T = 0$. The 
fact that we have a different time contour (the Keldysh contour C as opposed to the usual time axis) does not 
change the calculation, for the integrals involved in Eq.\ (\ref{boss}) are over space coordinate and the bose
fields commute no matter which branch of the time contour they are on. Another way of saying this is that the 
Keldysh operators on different branches still correspond to the same particle. Dividing the time contour in two
parts is a matter of mathematical convenience rather than physical distinction.

The cut-off 
$\alpha$ is a lattice spacing and so should be the same for both bosons and fermions. The operator equality Eq.\ (\ref{opereq})
means that any correlation function (in the limit $\alpha\rightarrow 0$), 
calculated in the fermi theory with the cut-off $\alpha$, is reproduced
in the bosonic theory with the same cut-off if the fermion operator in the left-hand side is replaced by the 
bosonic operator in the right-hand side of Eq.\ (\ref{opereq}). 
Using this operator equivalence we can construct the boson Hamiltonian from the fermion Hamiltonian Eq.\ (\ref{fermham}).
The fermion kinetic energy corresponds
to that of bosons. The umklapp interaction term gives rise to the cosine interaction of the boson field. The
conjugate operators $P_{+}$ and $P_{-}$ cancel out, so the interaction is a function only of the boson fields
$\phi_{+}$ and $\phi_{-}$ itself, as it is in the zero-temperature bosonisation. The interaction constant $V$ now 
acquires the factor $1\over {\displaystyle \alpha^2}$ from the prefactor in Eq.\ (\ref{opereq}). The boson Hamiltonian
therefore is

\ 

\begin{equation}
H_B = (\partial_\mu \phi)^2 + V' \cos 2\beta\phi.
\end{equation}

\ 
 
\noindent
where $V'={{\displaystyle V}\over{\displaystyle \alpha^2}}$.

Due to the special form of the interaction ($\cos 2\beta\phi$) it is most convenient to formulate the Keldysh technique
in the path integral representation, developed by A. Schmid \cite{schm}. The boson action for our model
is

\ 

\begin{equation}
{\it A} = \int dx dt \left[ \phi^{*}_{\gamma} {\it B}_{\gamma\nu}^{\;} \phi_{\nu}^{\;} 
- V' ( \cos 2\beta\phi_{+}^{\;} - 
\cos 2\beta\phi_{-}^{\;} ) \right] \; ,
\end{equation}

\  

\noindent
where $\phi_{+}^{\;}$ and $\phi_{-}^{\;}$ are the two components of the Keldysh boson field, $\gamma$ and $\nu$ denote
Keldysh indices. 
${\it B}_{\mu\nu}^{\;}$ is the kinetic operator, so that the inverse  ${\it B}_{\mu\nu}^{-1}$ is proportional to 
the boson Green's function in the Keldysh basis

\  

\begin{eqnarray}
{\it B}_{\mu\nu}^{-1} = - i
\pmatrix
{
{\it D}^{++}&{\it D}^{+-}\cr
{\it D}^{-+}&{\it D}^{--}\cr
}
\end{eqnarray}

\  

\noindent
We shall later use these functions in the space-time representation (see Appendix A). 

The fermion density in the
conventional bosonisation is represented by the spatial derivative of the boson field $\varphi$

\ 

\begin{equation}
\rho={1\over{\sqrt\pi}} \partial_1 \varphi .
\nonumber
\end{equation}

\ 

\noindent
Here in the same manner we write the fermion density operators. The density-density correlators
are represented by functional integrals in which the predexponential is some linear combination of 
the spatial derivatives of the bose fields corresponding to that particular correlator. The 
advanced correlator is

\  

\begin{equation}
{\langle \rho \rho \rangle}_A = {\it \Pi} (x_1-x_2) = {1\over {\it Z}} \int {\cal D} \left[ \phi_1 , \phi_2 \right]
 \; \partial_1 \phi^{*}_{\alpha}(x_1) \tau_{\alpha\beta} \partial_1 \phi^{\;}_{\beta}(x_2) \; \exp({-{\it A}}) ,
\label{denint}
\end{equation}

\  

\noindent
where the density vertex $\tau_{\alpha\beta}$ is

\ 

\begin{equation}
\tau_{\alpha\beta} = {1\over\pi} 
\pmatrix
{
{1}&{-1}\cr
{1}&{-1}\cr
}.
\end{equation}

\  

We now check, following the approach of R. Shankar\cite{shan}, that the above construction yields correct 
results in perturbation theory.
The density-density correlation function for non-interacting fermions should be calculated 
separately for the "left" and "right" movers and the results should be added. That gives for 
the advanced correlator

\  

\ 

\begin{equation}
\Pi_0={2\over\pi} \; {q^2\over{\omega^2-q^2-2i\delta\omega}} \; ,
\label{fermnolorder}
\end{equation}

\  

\ 

\noindent
where $\delta$ is infinitesimately small. In the next order we have two topologically nontrivial diagrams given on Fig. 2
Again we have to repeat the calculation for the "left" movers and add the 
results with proper combinatorial factors. For simplicity we give here only the imaginary part of the first-order
correction, which for us is most important 

\  

\ 

\begin{equation}
{\it Im} \Pi_1 = {V^2\over{2\pi^2}} \; {q^2\over{\omega^2-q^2}} \; {\omega T\over{\omega^2-q^2}} .
\label{firstorderim}
\end{equation}

\  

\ 

\noindent
The next order of perturbation is discussed in the next Section.

We now turn to the calculation of the same perturbation series on the boson language.
In the non-interacting case we have to calculate a simple Gaussian integral which in momentum space gives

\ 

\begin{equation}
{\it \Pi}_0(\omega, q) = {2\over\pi} \; q^2  \; {\it D}_{A} (\omega , q) = 
{2\over\pi} \; {q^2\over{\omega^2-q^2-2i\delta\omega}}
 \; ,
\label{bosmnolorder}
\end{equation}

\ 

\ 

\noindent
which is the same as Eq.\ (\ref{fermnolorder}). ${\it D}_{A} (\omega , q)$ is the advanced boson Green's function,
connected to the original Keldysh basis via

\

\begin{equation}
{\it D}_{A} = {\it D}^{--} - {\it D}^{+-} .
\nonumber
\end{equation}

\ 

We now expand the exponent in the integral Eq.\ (\ref{denint}) in the series in $V'$. In the first non-trivial 
order we get 

\  

\begin{eqnarray}
{\it \Pi}_1(x_1-x_2) = &&{V'}^2 \; \int d^2y_1 d^2y_2 \; {1\over {\it Z}} 
\int {\cal D} \left[ \phi_1 , \phi_2 \right] \; 
\partial_1 \phi^{*}_{\alpha}(x_1) \tau_{\alpha\beta} \partial_1 \phi^{\;}_{\beta}(x_2) \;
\nonumber\\
&&\;
\nonumber\\
&&{ \ \ \ \ \ \ \ \ \ \ \ \ \ \ \ \ \ \ \ \ \ \ \ \ \ \ \ }
\exp\left(-\int dx dt \; \phi^{*}_{\mu} {\it B}_{\mu\nu}^{\;} \phi_{\nu}^{\;} \right)
\nonumber\\
&&\;
\nonumber\\
&&\left[ \cos 2\beta\phi_{+}^{\;}(y_1) - \cos 2\beta\phi_{-}^{\;}(y_1) \right] \; 
\left[ \cos 2\beta\phi_{+}^{\;}(y_2) - \cos 2\beta\phi_{-}^{\;}(y_2) \right] \; .
\label{firstorderint}
\end{eqnarray}

\  

\noindent
Since the cosines contribute the linear terms in the exponent, the integral is still Gaussian with the same
prefactor of the exponent. 
This feature will remain in all higher
orders of the perturbation series. The calculation of that integral we shall discuss in more detail due to its
importance for the later arguments.
In the first-order integral Eq.\ (\ref{firstorderint}) we have four different terms of the same type, arising
from the cosines. In 
the exponent they have combinations like $2\beta(\phi^{\;}_1(y_2)-\phi^{\;}_2(y_1))$ with all possible permutations of indices.
To calculate the functional integral we make the Fourier transform of the bose fields. We perform the transform
in the most general way, since the same expressions will appear later.
As we show below in any order of the perturbation theory we shall need to
calculate averages of the form

\ 

\begin{equation}
\langle \partial_1 \phi^{*}_{\alpha}(x_1) \tau_{\alpha\beta} \partial_1 \phi^{\;}_{\beta}(x_2) \;
\exp \left(\int {{d^2k}\over{{(2\pi)}^2}} \; {\; ^{({\bf r})}I^{*}_{\mu}}(k,y_i) \; 
\phi_{\mu}^{\;}(k) + {\it h. c.} \; \right) \rangle.
\label{aver}
\end{equation}

\ 

\noindent
For the first-order integral Eq.\ (\ref{firstorderint}) we four different factors ${\; ^{({\bf r})}I^{*}_{\mu}}(k)$:

\newpage 

\begin{eqnarray}
\; &&\; { \ \ \ \ \ \ \ \ \ }
\pmatrix
{
{ {\; ^{({\bf 1})}I^{*}_{1}}(k, y_i)}\cr
{ {\; ^{({\bf 1})}I^{*}_{2}}(k, y_i)}\cr
}=
\pmatrix
{
{i\beta e^{iky_1}}\cr
{-i\beta e^{iky_2}}\cr
}, 
\pmatrix
{
{ {\; ^{({\bf 2})}I^{*}_{1}}(k, y_i)}\cr
{ {\; ^{({\bf 2})}I^{*}_{2}}(k, y_i)}\cr
}=
\pmatrix
{
{-i\beta e^{iky_2}}\cr
{i\beta e^{iky_1}}\cr
},\;
\nonumber\\
&&\;
\nonumber\\
&&\;
\nonumber\\
&&\;
\pmatrix
{
{ {\; ^{({\bf 3})}I^{*}_{1}}(k, y_i)}\cr
{ {\; ^{({\bf 3})}I^{*}_{2}}(k, y_i)}\cr
}=
\pmatrix
{
{i\beta (e^{iky_1}-e^{iky_2})}\cr
{0}\cr
},
\pmatrix
{
{ {\; ^{({\bf 4})}I^{*}_{1}}(k, y_i)}\cr
{ {\; ^{({\bf 4})}I^{*}_{2}}(k, y_i)}\cr
}=
\pmatrix
{
{0}\cr
{i\beta (e^{iky_1}-e^{iky_2})}\cr
}.
\label{fofac}
\end{eqnarray}

\ 

\noindent
Completing the square in the exponent we calculate the 
functional integral in Eq.\ (\ref{firstorderint}) and
apart from the numerical factor we get in the momentum space

\  

\begin{eqnarray}
{\it \Pi}_1(\omega, q) \sim \int d^2k \; kq \int \prod d^2y_i \sum_{ \{ {\bf r} \} }
\; {\; ^{({\bf r})}I^{*}_{\mu}}(q, x_i) {\it B}_{\mu\alpha}^{-1}(q) \;
\tau_{\alpha\beta} \; {\it B}_{\beta\nu}^{-1}(k) {\; ^{({\bf r})}I^{\;}_{\nu}}(k, x_i)  K_1(y_i) ,
\label{spint}
\end{eqnarray}

\ 

\noindent
where 

\ 

\begin{eqnarray}
K_1(y_i)=\exp\left( \int {{d^2k'}\over{{(2\pi)}^2}} \; {\; ^{({\bf r})}I^{*}_{\gamma}}(k', y_i) {\it B}_{\gamma\delta}^{-1}(k')
{\; ^{({\bf r})}I^{*}_{\delta}}(k', y_i) \; \right) , 
\label{kern}
\end{eqnarray}

\  

\noindent
and the sum is over four different factors ${\; ^{({\bf r})}I^{\;}_{\nu}}(k, y_i)$ corresponding to the four different terms
of the type (\ref{aver}) in the first order integral Eq.\ (\ref{firstorderint}).

Here the integration measure $\prod d^2y_i$ is equal $dy_1dt_1dy_2dt_2$. All dependence on these variables is
contained in the factors ${\; ^{({\bf r})}I^{*}_{\mu}}(k, y_i)$. 
These factors are nothing but exponentials $\beta e^{iky}$ with 
different signs. Therefore the integral in the exponent in Eq.\ (\ref{kern}) is just the Fourier transform 
of the boson Green's function back from momentum space to the real space. The boson functions in real space
are logarithms, so all four exponents $K_1(x_i)$ can be evaluated. All four have the same structure
\begin{equation}
K_1(y=y_1-y_2) = { {f(y,t)} \over {S(y,t)} },
\label{firstor}
\end{equation}

\noindent
where 

\begin{equation}
S^{-1}(y,t) = { { {(\pi T \alpha)}^4 } \over { \sinh^2 \pi T (y-t) \; \sinh^2 \pi T (y+t) } } ,
\label{kernel}
\end{equation}

\  

\noindent
while $f(y,t)$ is some algebraic function, which has no poles. In
the long-time asymptotic $K_1(y, t)$ acts like a derivative of a delta-function. It has {(due to the
denominator Eq.\ (\ref{kernel}))} a sharp singularity
at $y, t = 0$ and decays very fast as the variables go away from that point. Therefore the integral over
$y, t$ is dominated by the small region of the size $\alpha$ around the origin. The remaining spatial integration over the
sum $y_1 + y_2$ involves only functions ${\; ^{({\bf r})}I^{*}_{\mu}}(k, y_i)$ from the prefactor since $K_1(y_i)$ depends only 
on the difference of the variables $y_1$ and $y_2$. Since ${\; ^{({\bf r})}I^{*}_{\mu}}(k, y_i)$ are exponentials the integral
is easily evaluated to yield $\delta (k-q)$, which solves the momentum integration. 
So after the summation over all four terms we get the result Eq.\ (\ref{firstorderim}), as expected.

The next orders of the perturbation series, which we discuss in the following Section, could be calculated
in the same manner and are also the same for the boson and the fermion formulations of he theory. The
bosonisation procedure is thus justified.

\ 

\section{Absence of Diffusion for the Heisenberg model}
\label{sec-res}

\ 

\ 

Although the boson and the fermion versions of the theory are completely equivalent and should give the 
same results in the perturbation theory, the bosonised version allows easier evaluation or the higher
orders of the perturbation theory. Indeed, all the higher order corrections differ from the first order
Eq.\ (\ref{firstorderint}) only by appearance of additional cosine brackets (and corresponding spatial
integrations). That means that in any order of perturbation we have to calculate averages of the form
Eq.\ (\ref{aver}). The functions ${\; ^{({\bf r})}I_{\mu}}(k, x_i)$ will be now sums of exponentials, for example in the next
order of perturbation we will have terms like

\ 

\begin{equation}
{\; ^{({\bf r})}I_{\mu}}(k, x_i)=\beta\left[e^{ikx_1}-e^{ikx_2}+e^{ikx_3}-e^{ikx_4}\right],
\end{equation}

\  

\noindent
similar to the first order factors Eq.\ (\ref{fofac}), but constructed from four different exponentials.
We can still perform the functional integration and get the same general formula Eq.\ (\ref{spint}). Since 
functions ${\; ^{({\bf r})}I_{\mu}}(k, x_i)$ are exponentials the integral in the exponent in Eq.\ (\ref{spint}) still
yields the boson Green's functions in the space-time representation, but now instead of one such a 
function as in the first order we have a sum of them. In the second order we shall get

\  

\begin{equation}
K_2(x_1, ... x_4) = { {S(x_2-x_3) S(x_1-x_4)} \over {S(x_1-x_2) S(x_1-x_3) S(x_2-x_4) S(x_3-x_4)} }\;
f_2(x_1, ... x_4),
\label{secondkernel}
\end{equation}

\  

\noindent
where $S(x)$ is the "singular" denominator, defined in Eq.\ (\ref{kernel}). 

In higher orders functions $K$
have the same structure, but with more factors $S(x)$ in the numerator and denominator. These factors
have the same singular behaviour as described in the previous Section. Therefore the integration over 
the multidimensional space will be dominated by the regions where two pairs of variables are almost 
equal to each other, for example $x_1=x_2$ and $x_3=x_4$. The integration over one such pair effectively
reduces the function Eq.\ (\ref{secondkernel}) to the form of the previous order Eq.\ (\ref{firstor}) (in
the long-time asymptotic). One can see it by inspecting Eq.\ (\ref{secondkernel}). 
Consider, for instance, the contribution from the region $x_1=x_2$. In this region 
$K_2 \sim S^{-1}(x_1-x_2)S^{-1}(x_3-x_4)$, so the integral over $x_2$ leads to the same form
($\sim 1/S(x_3-x_4)$), which is the main part of the first order function $K_1$ Eq.\ (\ref{firstor}). The integration
over $x_3$ and $x_4$ is then the same as in the first order .
Therefore the spatial integration in
in the higher order terms does not yield any additional singularity to that produced by the prefactor of the 
exponent, which is the same as in the first order Eq.\ (\ref{spint}) and contains just 
two boson Green's functions. So the structure of 
any term in the perturbation series is 

\ 

\begin{equation}
{\it Im} {\it \Pi}_1(\omega, q) \sim {\it P}{ {\omega q^2} \over { {(\omega^2-q^2)}^2 } } \; g(\omega, q).
\label{corpat}
\end{equation}

\ 

\noindent
Here $g(\omega, q)$ is some function, which has no poles at small $\omega, q$. We note that ${\it Im} {\it \Pi}_1(\omega, q)$
differs from the first order Eq.\ (\ref{firstorderint}) only by $g(\omega, q)$, and ${\it P}$ denotes principal value. 

Combining our conclusions we can write down the structure of the density-density correlation
function in our model.

\  

\begin{equation}
{\it \Pi}(\omega, q) = {2\over\pi} \; {q^2\over{\omega^2-q^2+2i\delta\omega}} \;
+i \; { {\omega T q^2} \over { {(\omega^2-q^2)}^2 } } \; g(\omega, q, V)
\label{result1}
\end{equation}

\  

\noindent
The significance of this result for us is in the fact, that the higher order corrections do not acquire 
additional poles in the imaginary part, which could sum up in a diffusive pole in ${\it \Pi}(\omega, q)$.

The fermion perturbation theory gives the same results in the low orders of the perturbation.
In the second non-trivial order we have six topologically  different
diagrams, presented on Fig. 3.  Our purpose is to show that the correction to the density-density
correlation function, which is the sum of these diagrams, is not qualitatively different from the 
first order result. The only
diagrams that produce the extra pole are the diagrams ${\it a}$ and ${\it b}$ on Fig. 3. They cancel each 
other exactly, so that the second order correction has the pole of the same order as the first order one.
This cancellation seems to be an accidental property of the problem in the fermion representation, but
the bosonisation approach shows that it happens 
in all orders of perturbation theory. 

Our results clearly show the absence of spin diffusion in the Heisenberg model. 
Instead of the diffusive pole in the density-density correlation function we 
found some kind of a propagating behaviour. 
Note that the second order pole in Eq.\ (\ref{result1}) should be regarded as a principal value, so the
imaginary part of general susceptibilities does not contain unphysical contributions proportional to
$\delta^2(\omega^2-q^2)$.
This result should have been expected.
In the boson representation our problem is essentially the Sine-Gordon model. It is known in the theory
of the Sine-Gordon equation that due to the infinite number of conserved charges the excitations of the
model are propagating and {\it not} diffusive. So by showing the mapping of the Heisenberg model onto the
Sine-Gordon model we have showed the absence of spin diffusion in the model Eq.\ (\ref{heisham}).

\ 

\section{Spin-phonon interaction}

\ 

\ 

The Heisenberg model Eq.\ (\ref{heisham}) does not exactly describes the physics of a real material. We now try to make it 
a bit more realistic. Even in the absence of disorder, at finite temperatures there always
are phonons in the system. The exchange integral depends in general on the instantaneous separation of magnetic
ions. We consider the linear approximation for spin-phonon coupling. That is we expand the separation-dependent
Heisenberg coupling constant to the first order in ionic displacement. $u(R_i)$

\ 

\begin{equation}
J(r_i-r_j)=J_0(R_i-R_j) + \chi \left ( u(R_i)-u(R_j) \right ),
\label{exi}
\end{equation}

\ 

\noindent
where $R_i$ is the equilibrium position of the ion and $r_i=R_i+u(R_i)$.  That leads to the simplest form of
the spin-phonon interaction Hamiltonian

\ 

\begin{equation}
{\cal H}_{\it sp-ph} = J' \sum_{<ij>} \vec S_i \vec S_j \; \left(b^{\dag}_i + b^{\;}_i\right).
\label{spphham}
\end{equation}

\ 

We shall now try, treating interaction Eq.\ (\ref{spphham}) as a small perturbation of the original Hamiltonian
Eq.\ (\ref{heisham}), to examine it's impact on the long-time spin dynamics. To do that we first bosonise
Eq.\ (\ref{spphham}). We get

\ 

\begin{equation}
{\cal H}_{\it b-ph} =  J'\sum_{k} {|\phi_k|}^2 \; \left(b^{\dag}_k + b^{\;}_k\right).
\end{equation}

\ 

\noindent
The boson self-energy in the first non-trivial order in ${\cal H}_{\it b-ph}$ is presented by the diagram on 
Fig. 4. The solid line represents a boson and the dashed - a phonon. For the  imaginary part we are interested
in we get

\  

\begin{eqnarray}
{\it Im} \Sigma (\omega, q) \; = \; \left\{
\begin{array}{ll}
\eta \omega  q v_F & \; , \; \; if \; \; q>T \\
\eta \omega  T & \; , \; \; if \; \; q<T \\
\end{array}
\right.
\label{selfen}
\end{eqnarray}

\  

\noindent
where restoring the units $\eta \sim {(J'/c)}^2$ and $c$ is the speed of sound. The momentum dependence 
of ${\it Im} \Sigma (\omega, q)$ arises due to the momentum 
in the numerator of the phonon Green's function

\  

\begin{equation}
{\cal D}_{ph}(\omega, q) = { c^2 q^2 \over {\omega^2 - c^2 q^2 +i\delta\omega}},
\end{equation}

\  

\noindent
where $\delta$ is infinitesimally small.
Note that although we consider 1D spin chain, the phonons in a real material are three-dimensional, so when evaluating
the self-energy Eq.\ (\ref{selfen}) one must integrate out the two other components of the phonon momentum.

We shall now investigate, how the umklapp term renormalizes this self-energy. We now divide the boson field
into "slow" and "fast" parts, separated by some cut-off $k_0$. We integrate out all "fast" degrees of 
freedom (those with momentum bigger the cut-off) and see how the imaginary part of the boson self-energy 
(namely the coefficient $\eta$) changes with the cut-off. The result is that when the cut-off is bigger 
than the temperature, $T$,  $\eta$ rises as some power of the cut-off. But after the cut-off becomes smaller than the 
temperature, the imaginary part grows exponentially. That means, that on the large distances the motion becomes 
diffusive. We notice that it should happen in any experimental situation, because it is just the 
existence of the spin-phonon interaction (no matter how small) is responsible for the diffusion.

After we introduce the "slow" and "fast" variables as

\

\begin{equation}
\phi_\mu^{\;}=\phi_\mu^{\it s} + \phi_\mu^{\it f}.
\end{equation}

\

\noindent
The integral Eq.\ (\ref{denint}) then is

\ 

\begin{equation}
{\it \Pi} (x_1-x_2) = {1\over {\it Z}} \int {\cal D} \left[ \phi_1^{\it s}, \phi_2^{\it s} \right]
{\cal D} \left[ \phi_1^{\it f}, \phi_2^{\it f} \right]
 \; \partial_1 \phi^{{\it s}*}_{\alpha}(x_1) \tau_{\alpha\beta} \partial_1 \phi^{\it s}_{\beta}(x_2) 
\; \exp({-{\it A}}) ,
\label{dintfs}
\end{equation}

\  

\noindent
where the prefactor contains only slow degrees of freedom since we are looking for the infrared asymptotic.
We can separate the integral over fast variables and the density integral now becomes

\ 

\begin{eqnarray}
{\it \Pi} (x_1-x_2) = {1\over {\it Z'}} \int {\cal D} \left[ \phi_1^{\it s}, \phi_2^{\it s} \right] &&\;
\partial_1 \phi^{{\it s}*}_{\alpha}(x_1) \tau_{\alpha\beta} \partial_1 \phi^{\it s}_{\beta}(x_2) \;
\nonumber\\
&& \;
\nonumber\\
&&\exp \left( \int {{d^2k'}\over{{(2\pi)}^2}} \left[ \phi^{s*}_{\mu} {\it B}_{\mu\nu}^{\;} \phi_{\nu}^{s}
\right] \right) \;
\exp \left( \ln M \right),
\label{fi}
\end{eqnarray}

\ 

\noindent
where $M$ is the integral over the fast variables, which in the first non-trivial order in perturbation is a 
Gaussian integral without any prefactor

\ 

\begin{eqnarray}
M= &&{ {V'^2}\over2 } \int {\cal D} \left[ \phi_1^{\it f}, \phi_2^{\it f} \right] \;
\exp \left( \int {{d^2k'}\over{{(2\pi)}^2}} \left[ \phi^{f*}_{\mu} {\it B}_{\mu\nu}^{\;} \phi_{\nu}^{f} \right] \right)
\int d^2x_1 d^2x_2
\nonumber\\
&& \;
\nonumber\\
&& 
\Bigg[ \cos 2\beta \bigg(\phi_1^{s}(x_1)+\phi_1^{f}(x_1)\bigg) - \cos 2\beta\bigg(\phi_2^{s}(x_1)+\phi_2^{f}(x_1)\bigg)\Bigg] 
\nonumber\\
&& \;
\nonumber\\
&& 
\Bigg[ \cos 2\beta \bigg(\phi_1^{s}(x_2)+\phi_1^{f}(x_2)\bigg) - \cos 2\beta\bigg(\phi_2^{s}(x_2)+\phi_2^{f}(x_2)\bigg)\Bigg]. 
\end{eqnarray}

\  

\noindent
In functional integral Eq.\ (\ref{fi})
$\ln M$ plays a role of renormalisation of the imaginary part of the kinetic operator.
Now we calculate the functional integral and expand the result in the boson fields $\phi_i^{s}$ to get their bilinear
combination, which gives the renormalisation of $\eta$. We get

\

\begin{equation}
M=  { {{(V'\beta)}^2}\over4 }\int d^2x_1 d^2x_2 
\Bigg[ \phi_\alpha^{s}(x_1) {\cal R}_{\alpha\beta}(x_{12})\phi_\beta^{s}(x_2) \Bigg],
\label{renormint}
\end{equation}

\ 

\noindent
where as usual $x_{12}=x_1-x_2$. The elements of the matrix ${\cal R}_{\alpha\beta}(x)$ are the exponentials of the 
boson Green's functions in the space-time representation. They are presented in Appendix B, where we give the
detailed renormalisation calculation.
The boson Green's functions entering the matrix ${\cal R}_{\alpha\beta}(x_{12})$
are now different from those in Appendix A, due to the self-energy Eq.\ (\ref{selfen}).
The retarded function in the momentum space is now

\ 

\begin{equation}
{\it D}_R(\omega,q)={1\over{\omega^2-q^2+i{\it Im} \Sigma (\omega, q)}}
\end{equation}

\ 

\noindent
and the corresponding functions ${\it D}_A(\omega,q)$ and ${\it D}_F(\omega,q)$ acquire the same self-energy.
Note, that ${\cal R}_{\alpha\beta}(x_{12})$ is expressed via boson Green's functions in the not rotated Keldysh
basis. However, 
the coefficient $\eta$ is easier to extract from the Green's functions in the rotated Keldysh basis Eq.\ (\ref{rotbas}). 
The element ${\cal R}_{R}$ of the matrix ${\cal R}_{\alpha\beta}$ in the rotated basis which corresponds
to the renormalisation of the retarded function is

\  

\begin{equation}
{\cal R}_R={1\over2} \bigg( {\cal R}_{11}(x)-{\cal R}_{22}(x)-{\cal R}_{12}(x)+{\cal R}_{21}(x)
\bigg).
\end{equation}

\  

\noindent
If the cut-off is large, $v_F k_0 > T$, this yields after the Fourier transform 

\ 

\begin{equation}
{\cal R}_R=i\omega {\displaystyle{ \eta }
\over{\displaystyle k_0^3}} \exp(-{{\displaystyle\beta^2}\over{\displaystyle 2\pi}} \ln k_0\alpha),
\end{equation}

\  

\noindent
so that the change in $\eta$ is

\  

\begin{equation}
\Delta\eta=\eta \; { V^2 \over {4{(k_0\alpha)}^6}},
\label{etaren}
\end{equation}

\  

\noindent
since $\beta=\sqrt{4\pi}$. Here as always, $\alpha$ is the lattice spacing which cuts-off the large momentum
integration.So here we have a power-law rise in $\eta$.
When the cut-off is less that the temperature we get exponential renormalisation:

\  

\begin{equation}
\Delta\eta=\eta_T \; \exp \Bigg( 4 {T\over{v_F k_0}} \Bigg) \; 
{ V^2 \over {4{(k_0\alpha)}^4}}.
\label{etarez}
\end{equation}

\  

\noindent
Here $\eta_T$ is the effective damping at the scales $v_F k_0 \sim T$.
That means, that on the scales of momenta less that the temperature, the imaginary part coefficient $\eta$
grows very rapidly, and we immediately get the diffusive dynamics.
The renormalisation group treatment breaks down
when the correction  Eq.\ (\ref{etarez}) becomes of the order of unity. That determines the length scales, 
where the dynamics becomes diffusive. The mean-free path is 

\ 

\begin{equation}
{\it l} \sim {v_F\over T} \ln(\eta_0 V^2),
\end{equation}

\ 

\noindent
where $\eta_0$ is the original value of the coefficient $\eta$, proportional to the spin-phonon coupling
constant (Eq.\ (\ref{selfen})). 

The estimate of the mean-free path allows us to estimate the diffusion coefficient as $D={\it l} v_F$, so that

\ 

\begin{equation}
D\sim{v_F^2\over T} \ln({J'V\over c}).
\label{d}
\end{equation}

\

\section{Conclusions}

\ 

\ 

We now briefly review our results. We started from 1D Heisenberg Hamiltonian Eq.\ (\ref{heisham}). Out goal was to
calculate the spin-spin correlation function at long times and non-zero temperature.
and check whether or not it had a diffusive pole Eq.\ (\ref{corrfunc})
in some region in phase space. To perform such a calculation we mapped our original problem onto 1D fermion
model Eq.\ (\ref{fermham}) using the Jordan-Wigner transformation Eq.\ (\ref{jordwig}). We used the usual diagrammatic 
technique to calculate a few first orders of the perturbation theory (shown 
on Fig. 2 and Fig. 3). In order to go beyond that simple approximation we bosonised the fermion
model. To get the long-time asymptotic of the spin-spin correlator, we had to combine bosonisation with 
Keldysh technique, which allows one to calculate real-time correlation functions at finite temperatures
without any need in analytic continuation. 

The bosonisation procedure allowed us to find the general form of
higher order corrections and to sum up the perturbation series. It turned out that
in each order of perturbation the correction to the spin-spin correlator had the same form Eq.\ (\ref{corpat}),
therefore we concluded, that the exact spin-spin correlation function do not acquire the diffusive pole
from the summation of the perturbation series. This result
is also known in the theory of the Sine-Gordon model (which coincides with our boson model), where it has been 
found that due to the infinite number of conserved charges excitations are propagating and not diffusive.

We also check whether these results (the absence of spin diffusion in the perturbation theory) are robust with respect
to small dissipation effects present in real physical systems. Specifically, we considered the effect of a weak 
spin-phonon interaction Eq.\ (\ref{spphham}). We mapped the full problem (including the spin-phonon interaction)
to the bosonic model. We found that the interaction with phonons leads to the boson self-energy Eq.\ (\ref{selfen}),
the imaginary part of which is proportional to the constant, $\eta$, at small momenta. Further, we applied the 
renormalisation group analysis; we integrated out ``fast'' degrees of freedom and showed that this constant $\eta$
grows moderately while the cut-off is larger than temperature (Eq.\ (\ref{etaren})), but grows exponentially after the cut-off 
becomes smaller than the temperature (Eq.\ (\ref{etarez})). At scales where the imaginary part of the spin-spin correlator
becomes of the order of the real part the spin dynamics becomes purely diffusive. By associating the mean-free path 
with the scale on which the renormalisation procedure breaks down (namely, the
renormalisation Eq.\ (\ref{etarez}) becomes of the order of unity) we estimate the diffusion coefficient Eq.\ (\ref{d}).
Restoring the original units and estimating the spin-phonon coupling constant $J'$ from the expansion of the exchange 
integral Eq.\ (\ref{exi}), we have

\ 

\begin{equation}
D\sim{{\pi^2 {(J\alpha)}^2}\over {\hbar k_B T}} \ln({ {J \alpha} \over {\hbar c}}).
\end{equation}
 
\ 

\noindent
where $\alpha$ is the lattice spacing and $c$ is the speed of sound.

Thus we found that the presence of the 
spin-phonon interaction changes the long-time behaviour of the spin-spin correlator, which becomes diffusive.

\ 

\section*{Acknowledgments}

\ 

\ 

The author is greatly indebted to Prof. L. Ioffe for drawing his attention to the spin-diffusion problem and for most
stimulating discussions.

\ 

\section*{Appendix A}

\ 

\ 

We present here the boson and fermion Green's functions, which we are using in our calculations. It is the 
certain similarity between them, that had inspired the bosonisation. As in the usual zero-temperature
bosonisation we need the Green's functions in the space-time representations. We perform the bosonisation in 
the original Keldysh basis, but for simplicity we here calculate Green's functions in the rotated basis and
them transform them back.

The retarded fermion Green's function in the momentum space is 

\ 

\begin{equation}
{\it G}_R (\epsilon, k) = { 1\over {\epsilon \mp k +i0} }
\label{fermfunc}
\end{equation}

\ 

\noindent
where "-" is for the "left" and "+" for the "right" movers. For the simplification of the formulas the Fermi
velocity is set equal to unity. The Fourier integral, which one has to calculate in order to transform the
function Eq.\ (\ref{fermfunc}) to the real space is formally divergent at large momenta. As usual in 1D
calculations we introduce a cut-off $\alpha$ by adding the exponent $e^{-\alpha |k|}$ to the
integral. Thus we have

\  

\begin{equation}
{\it G}_R (x,t)=\int {{d\epsilon dk}\over{{(2\pi)}^2}} \; e^{ikx-i\epsilon t} \; { 1\over {\epsilon \mp k +i0} }
\; {e^{-\alpha|k|}} .
\label{furintf}
\end{equation}

\  

\noindent
Now we have the perfectly converging integral and get

\  

\begin{equation}
{\it G}_R (x,t)=-{ {\theta(t)} \over {2\pi} }  { {2i\alpha} \over { {(x \mp t)}^2+\alpha^2 } },
\end{equation}

\noindent
where 

\begin{eqnarray*}
\theta(t) \; = \; \left\{
\begin{array}{ll}
1 & \; , \; \; if \; \; t>0 \cr
0 & \; , \; \; if \; \; t<0 \cr
\end{array}
\right.
\end{eqnarray*}

\noindent
Similarly, the advanced function is

\  

\begin{equation}
{\it G}_A (x,t)={ {\theta(-t)} \over {2\pi} }  { {2i\alpha} \over { {(x \mp t)}^2+\alpha^2 } }
\end{equation}

\  

\noindent
The third Keldysh function in this basis for the right movers in momentum space is

\begin{equation}
{\it F}(\epsilon, k) = - 2 \pi i \tanh {k\over{2T}} \delta(\epsilon-k).
\label{ff}
\end{equation}

\noindent
The Fourier integral in converging and we get 

\  

\begin{equation}
{\it F}(x,t)= { T \over {\sinh \pi T (x-t)}}.
\label{fermf}
\end{equation}

\  

\noindent
The function for the left movers has the opposite sign of the time variable. To be more careful with the
pole one has to add $i\alpha$ to the space coordinate in Eq.\ (\ref{fermf}).

That completes the calculation of the fermions Green's functions in the rotated Keldysh basis

\begin{equation}
{\it G}_{\it rot}=
\pmatrix
{
0&{\it G}_{R}\cr
{\it G}_{A}&{\it F}\cr
}.
\label{rotbas}
\end{equation}

\ 

\noindent
To return to the original basis, which is needed for bosonisation, one has to perform the rotation

\begin{equation}
{\it G} = 
\pmatrix
{
{\it G}^{++}&{\it G}^{+-}\cr
{\it G}^{-+}&{\it G}^{--}\cr
}
= {\it R} {\it G}_{\it rot} {\it R}^{-1},
\label{orbas}
\end{equation}

\

\noindent
where the rotation matrix is given by

\  

\begin{equation}
{\it R}={ 1\over{\sqrt2}} \;
\pmatrix
{
1&1\cr
-1&1\cr
}.
\end{equation}

\ 

We now calculate the boson functions. We start again with the rotated basis. The retarded Green's function
in the momentum space is

\  

\begin{equation}
{\it D}_R(\omega, q)={1\over{\omega^2-q^2+2i\delta\omega}}.
\end{equation}

\  

\noindent
Again we have to introduce the cut-off $\alpha$. It is essential for the purposes of bosonisation to do it
in exactly the same way as for the case of fermions,  Eq.\ (\ref{furintf}). That way we get 

\  

\begin{equation}
{\it D}_R(x,t)=-{ i\over {4\pi} } \; {\theta(t)} \;
\ln { {(x-t+i\alpha)(x+t-i\alpha)} \over {(x-t-i\alpha)(x+t+i\alpha)} }.
\end{equation}

\  

\noindent
For the advanced function we get the same logarithm

\  

\begin{equation}
{\it D}_A(x,t)={ i\over {4\pi} } \; {\theta(-t)} \;
\ln { {(x-t+i\alpha)(x+t-i\alpha)} \over {(x-t-i\alpha)(x+t+i\alpha)} }.
\end{equation}

\  

\noindent
The third function in this basis, ${\it D}_{F}$ contains delta function as it's fermion counterpart Eq.\ (\ref{ff})

\begin{equation}
{\it D}_{F}(\omega, q) = - { {i \pi} \over {\omega} } \; \coth {\omega \over {2T}}  \;
\left( \delta(\omega-q) + \delta(\omega+q) \right),
\end{equation}

\  

\noindent
but we have to introduce the cut-off here. In real space we get

\  

\begin{equation}
{\it D}_{F}(x,t)={ i\over {2\pi} } \; \ln { { {(\pi T \alpha)}^2 } \over { \sinh \pi T(x-t) \sinh \pi T(x+t) } },
\end{equation}

\  

\noindent
where the zeros of the denominator should be treated in exactly the same way as in the fermion case ( see text
after Eq.\ (\ref{fermf}) ).

One can clearly see, that the fermion functions and the arguments of the logarithm in the boson functions are
constructed from the same elements. That is why the bosonisation works. To get exactly the same results in
the fermion and boson perturbation series we have to turn to the original Keldysh basis Eq.\ (\ref{orbas})
and calculate physical quantities, like the density-density correlation function, which is independent of
Keldysh indices and therefore of the choice of the calculational technique. Then the boson and the fermion versions
are the same in the limit $\alpha=0$ (the physical quantities should not depend on that cut-off).

\section*{Appendix B}

We start from the functional integral for the density-density correlation function Eq.\ (\ref{denint}),
where the kinetic term contains now non-zero imaginary part Eq.\ ({selfen}). We
are interested in the infrared asymptotic of that correlator. Therefore we divide the boson field $\phi_\mu$
in two part - "fast" and "slow"

\begin{equation}
\phi_\mu^{\;}=\phi_\mu^{\it s} + \phi_\mu^{\it f}.
\end{equation}

\noindent
The integral Eq.\ (\ref{denint}) then becomes

\begin{equation}
{\it \Pi} (x_1-x_2) = {1\over {\it Z}} \int {\cal D} \left[ \phi_1^{\it s}, \phi_2^{\it s} \right]
{\cal D} \left[ \phi_1^{\it f}, \phi_2^{\it f} \right]
 \; \partial_1 \phi^{{\it s}*}_{\alpha}(x_1) \tau_{\alpha\beta} \partial_1 \phi^{\it s}_{\beta}(x_2) 
\; \exp({-{\it A}}) ,
\label{dintfsB}
\end{equation}

\  

\noindent
where the prefactor contains only slow degrees of freedom since we are looking for the infrared asymptotic.
We can separate the integral over fast variables as

\begin{eqnarray}
I={1\over {\it Z''}} \int {\cal D} &&\left[ \phi_1^{\it f}, \phi_2^{\it f} \right] \;
\exp \left( \int {{d^2k'}\over{{(2\pi)}^2}} \left[ \phi^{f*}_{\mu} {\it B}_{\mu\nu}^{\;} \phi_{\nu}^{f} \right. \right.
\nonumber\\
&& \;
\nonumber\\
&&
- V' ( \cos 2\beta\phi_1^{s} \cos 2\beta\phi_1^{f} - \sin 2\beta\phi_1^{s} \sin 2\beta\phi_1^{f}
- \cos 2\beta\phi_2^{s} \cos 2\beta\phi_2^{f}
\nonumber\\
&& \left.\left. { \ \ \ \ \ \ \ \ \ \ \ \ \ \ \ \ \ \ \ \  \ \ \ }
  + \sin 2\beta\phi_2^{s} \sin 2\beta\phi_2^{f}) \right]
\right).
\label{fastint}
\end{eqnarray}

\  

\noindent
The density integral becomes now

\begin{eqnarray}
{\it \Pi} (x_1-x_2) = {1\over {\it Z'}} \int {\cal D} \left[ \phi_1^{\it s}, \phi_2^{\it s} \right] &&\;
\partial_1 \phi^{{\it s}*}_{\alpha}(x_1) \tau_{\alpha\beta} \partial_1 \phi^{\it s}_{\beta}(x_2) \;
\nonumber\\
&& \;
\nonumber\\
&&\exp \left( \int {{d^2k'}\over{{(2\pi)}^2}} \left[ \phi^{s*}_{\mu} {\it B}_{\mu\nu}^{\;} \phi_{\nu}^{s}
\right] \right) \;
\exp \left( \ln M \right),
\end{eqnarray}

\  

\noindent
where $\ln M$ plays now a role of renormalisation of the imaginary part of the kinetic operator.

We now calculate the integral over the fast degrees of freedom Eq.\ (\ref{fastint}). We expand the exponent
in Eq.\ (\ref{fastint}) up to the first non-trivial order in perturbation (which is actually all we need,
noting the results of Section~\ref{sec-res}) and get the Gaussian integral without any prefactor 

\begin{eqnarray}
M= &&{ {V'^2}\over2 } \int {\cal D} \left[ \phi_1^{\it f}, \phi_2^{\it f} \right] \;
\exp \left( \int {{d^2k'}\over{{(2\pi)}^2}} \left[ \phi^{f*}_{\mu} {\it B}_{\mu\nu}^{\;} \phi_{\nu}^{f} \right] \right)
\int d^2x_1 d^2x_2
\nonumber\\
&& \;
\nonumber\\
&& 
\left[ \cos 2\beta\phi_1^{s}(x_1) \cos 2\beta\phi_1^{f}(x_1) - \sin 2\beta\phi_1^{s}(x_1) \sin 2\beta\phi_1^{f}(x_1)
- \cos 2\beta\phi_2^{s}(x_1) \cos 2\beta\phi_2^{f}(x_1) \right.
\nonumber\\
&& \left. { \ \ \ \ \ \ \ \ \ \ \ \ \ \ \ \ \ \ \ \  \ \ \ }
  + \sin 2\beta\phi_2^{s}(x_1) \sin 2\beta\phi_2^{f}(x_1)) \right]
\nonumber\\
&& \;
\nonumber\\
&& 
\left[ \cos 2\beta\phi_1^{s}(x_2) \cos 2\beta\phi_1^{f}(x_2) - \sin 2\beta\phi_1^{s}(x_2) \sin 2\beta\phi_1^{f}(x_2)
- \cos 2\beta\phi_2^{s}(x_2) \cos 2\beta\phi_2^{f}(x_2) \right.
\nonumber\\
&& \left. { \ \ \ \ \ \ \ \ \ \ \ \ \ \ \ \ \ \ \ \  \ \ \ }
  + \sin 2\beta\phi_2^{s}(x_2) \sin 2\beta\phi_2^{f}(x_2)) \right].
\end{eqnarray}

\  

\noindent
We can now calculate the functional integral. The result is

\begin{eqnarray}
M= { {V'^2}\over4 }\int d^2x_1 d^2x_2 && \Bigg\{ 
\bigg[\cos 2\beta\phi_1^{s}(x_1)\cos 2\beta\phi_1^{s}(x_2)+\sin 2\beta\phi_1^{s}(x_1) sin 2\beta\phi_1^{s}(x_2) 
\bigg] \;
\nonumber\\
&& { \ \ \ \ \ \ \ \ }
\exp\left(i\beta^2\int_{k'>k_0} {{d^2k'}\over{{(2\pi)}^2}} {\it D}^{--}_k 
\left(1-e^{ikx}\right)\left(1-e^{-ikx}\right)\right)
\nonumber\\
&& \;
\nonumber\\
&&+ 
\bigg[\cos 2\beta\phi_2^{s}(x_1)\cos 2\beta\phi_2^{s}(x_2)+\sin 2\beta\phi_2^{s}(x_1) sin 2\beta\phi_2^{s}(x_2) 
\bigg] \;
\nonumber\\
&& { \ \ \ \ \ \ \ \ }
\exp\left(i\beta^2\int_{k'>k_0} {{d^2k'}\over{{(2\pi)}^2}} {\it D}^{++}_k 
\left(1-e^{ikx}\right)\left(1-e^{-ikx}\right)\right)
\nonumber\\
&& \;
\nonumber\\
&&- 
\bigg[\cos 2\beta\phi_1^{s}(x_1)\cos 2\beta\phi_2^{s}(x_2)+\sin 2\beta\phi_1^{s}(x_1) sin 2\beta\phi_2^{s}(x_2) 
\bigg] \;
\nonumber\\
&& { \ \ \ \  }
\exp\left(i\beta^2\int_{k'>k_0} {{d^2k'}\over{{(2\pi)}^2}} \left( {\it D}^{++}_k + {\it D}^{--}_k
-{\it D}^{-+}_ke^{ikx}-{\it D}^{+-}_ke^{-ikx}\right)\right)
\nonumber\\
&& \;
\nonumber\\
&&- 
\bigg[\cos 2\beta\phi_2^{s}(x_1)\cos 2\beta\phi_1^{s}(x_2)+\sin 2\beta\phi_2^{s}(x_1) sin 2\beta\phi_1^{s}(x_2) 
\bigg] \;
\nonumber\\
&& { \ \ \ \  }
\exp\left(i\beta^2\int_{k'>k_0} {{d^2k'}\over{{(2\pi)}^2}} \left( {\it D}^{++}_k + {\it D}^{--}_k
-{\it D}^{-+}_ke^{-ikx}-{\it D}^{+-}_ke^{ikx}\right)\right)
\Bigg\},
\label{renbigint}
\end{eqnarray}

\  

\noindent
where $x=x_1-x_2$ and $k_0$ is the momentum cut-off, delimiting fast variables from slow. 

The remaining integrals are similar to those, calculated in the regular perturbation series. We have the 
Fourier transform of the boson Green's functions to the real space in the exponent, then take the exponential
and Fourier transform back to the momentum space. The difference is that we now have another set of Green's
functions - with non-zero imaginary part - and the momentum integration in limited by the cut-off $k_0$.
The result of this integration will now depend on the cut-off. For the retarded and advanced functions we
get

\  

\begin{equation}
{\it D}_R(x, t; k_0)= { i\over {4\pi} } \; \theta(t) \; e^{-\eta' t} 2i {\it Im}
\Bigg\{ E_1\bigg[k_0 \left(\eta t-i(x+t)\right)\bigg] - E_1\bigg[k_0 \left(\eta t+i(x-t)\right)\bigg]\Bigg\},
\end{equation}

\begin{equation}
{\it D}_A(x, t; k_0)= - { i\over {4\pi} } \; \theta(-t) \; e^{-\eta' |t|} 2i {\it Im}
\Bigg\{ E_1\bigg[k_0 \left(\eta t-i(x+t)\right)\bigg] - E_1\bigg[k_0 \left(\eta t+i(x-t)\right)\bigg]\Bigg\},
\end{equation}

\  

\noindent
where $E_1$ is the exponential integral.
The result for the third Keldysh function ${\it D}_F$ depends on whether the cut-off is larger or smaller than
the temperature. For $k_0>T$ we get 

\  

\begin{equation}
{\it D}_F(x, t; k_0>T) = - { i\over {2\pi} } \; 
{1\over 2} {\it Re} \Bigg[ E_1\bigg[k_0 \left(\alpha +i(x+t)\right)\bigg] 
+ E_1\bigg[k_0 \left(\alpha +i(x-t)\right)\bigg]\Bigg]
\end{equation}

\  

\noindent
so that at the origin

\begin{equation}
{\it D}_F(0; k_0>T) = {i \over {2\pi}} \ln (k_0\alpha)
\end{equation}

\ 

\noindent
which is a small number.
For the other case, $k_0<T$, in the limit of large $x$ and $t$ and assuming that original $\eta$ is much
less than the temperature we have

\  

\begin{eqnarray}
{\it D}_F(x, t; k_0<T) = - {{iT}\over{\pi k_0}} \left( { {\sin k_o(x+t)} \over {k_o(x+t)} } + 
{ {\sin k_o(x-t)} \over {k_o(x-t)} } \right),
\end{eqnarray}

\  

\noindent
and at the origin is

\begin{equation}
{\it D}_F(0; k_0<T) = -{i \over \pi} { T \over k_0 },
\end{equation}

\ 

\noindent
which is extremely big. 

We can now proceed with the renormalisation of the imaginary part coefficient $\eta$. To do that we expand
Eq.\ (\ref{renbigint}) in boson fields and get their bilinear combination

\  

\begin{eqnarray}
M= && { {{(V'\beta)}^2}\over4 }\int d^2x_1 d^2x_2 
\nonumber\\
&& \;
\nonumber\\
&& \Bigg\{ 
\phi_1^{s}(x_1)\phi_1^{s}(x_2) \;
\exp\left(i\beta^2\int_{k'>k_0} {{d^2k'}\over{{(2\pi)}^2}} {\it D}^{--}_k 
\left(1-e^{ikx}\right)\left(1-e^{-ikx}\right)\right)
\nonumber\\
&& \;
\nonumber\\
&&+ 
\phi_2^{s}(x_1)\phi_2^{s}(x_2)
\exp\left(i\beta^2\int_{k'>k_0} {{d^2k'}\over{{(2\pi)}^2}} {\it D}^{++}_k 
\left(1-e^{ikx}\right)\left(1-e^{-ikx}\right)\right)
\nonumber\\
&& \;
\nonumber\\
&&- 
\phi_1^{s}(x_1)\phi_2^{s}(x_2)
\exp\left(i\beta^2\int_{k'>k_0} {{d^2k'}\over{{(2\pi)}^2}} \left( {\it D}^{++}_k + {\it D}^{--}_k
-{\it D}^{-+}_ke^{ikx}-{\it D}^{+-}_ke^{-ikx}\right)\right)
\nonumber\\
&& \;
\nonumber\\
&&- 
\phi_2^{s}(x_1)\phi_1^{s}(x_2)
\exp\left(i\beta^2\int_{k'>k_0} {{d^2k'}\over{{(2\pi)}^2}} \left( {\it D}^{++}_k + {\it D}^{--}_k
-{\it D}^{-+}_ke^{-ikx}-{\it D}^{+-}_ke^{ikx}\right)\right)
\Bigg\},
\label{rensmint}
\end{eqnarray}

\  

\noindent
which can be abbreviated as

\  

\begin{equation}
M =  { {{(V'\beta)}^2}\over4 }\int d^2x_1 d^2x_2 
\Bigg[ \phi_\alpha^{s}(x_1) {\cal R}_{\alpha\beta}(x)\phi_\beta^{s}(x_2) \Bigg],
\label{renormintB}
\end{equation}

\ 

\noindent
where as usual $x=x_1-x_2$. 

The imaginary part coefficient $\eta$ is most transparent in the rotated Keldysh basis Eq.\ (\ref{rotbas}). 
The element of the matrix ${\cal R}_{\alpha\beta}(x)$ in the rotated basis which corresponds
to the renormalisation of $\eta$ in the retarded function in terms of
original elements is

\  

\begin{equation}
{\cal R}_R={1\over2} \bigg( {\cal R}_{11}(x)-{\cal R}_{22}(x)-{\cal R}_{12}(x)+{\cal R}_{21}(x)
\bigg).
\end{equation}

\  

\noindent
For the case of the large cut-off this yields after the Fourier transform 
$i\omega \eta {\displaystyle{k_0}
\over{\displaystyle k_0^4}} \exp(-{{\displaystyle\beta^2}\over{\displaystyle 2\pi}} \ln k_0\alpha)$, so that the change in $\eta$ is

\  

\begin{equation}
\Delta\eta=\eta \; { V^2 \over {4{(k_0\alpha)}^6}},
\label{etarenB}
\end{equation}

\  

\noindent
since $\beta=\sqrt{4\pi}$. Here as always, $\alpha$ is the lattice spacing which cuts-off the large momentum
integration.So here we have a power-law rise in $\eta$.
When the cut-off is less that the temperature we gain different exponential, so that now

\  

\begin{equation}
\Delta\eta=\eta \; \exp \Bigg( 4 {T\over{k_0}} \Bigg) \; 
{ V^2 \over {4{(k_0\alpha)}^4}}.
\label{etarezB}
\end{equation}

\  

\noindent
That means, that on the scales of momenta less that the temperature, the imaginary part coefficient$\eta$
experiences a tremendous growth, and we immediately get the diffusive dynamics.

\end{document}